\documentclass[twoside]{article}

\usepackage{amsmath}
\usepackage{amssymb}
\usepackage{pstricks}
\usepackage{epsfig}
\usepackage{fleqn,espcrc2}

\newcommand{\msbar}{\overline{\rm MS}}

\title{An explicit example of HQET at one-loop order of perturbation theory}

\author{Martin Kurth\address{Department of Physics and Astronomy,
    University of Southampton, Highfield, Southampton,\\ SO17 1BJ, UK}
     and Rainer Sommer\address{DESY,
    Platanenallee 6, 15738 Zeuthen, Germany}}

\begin{document}

\begin{abstract}
As an explicit example of HQET, we construct correlation functions
containing a heavy-light axial current, both in the static approximation
and in full QCD. This enables us to investigate the size of finite mass
effects. As we use the lattice regularisation, we can also check the mass
dependence of discretisation errors.
\end{abstract}

\maketitle

\section{CORRELATION FUNCTIONS}

The axial current of one light and one heavy quark flavour is
phenomenologically interesting, as the decay constant $f_{\rm B}$ 
of the B-meson
is defined by a matrix element of such a current. This matrix element
requires a non-perturbative treatment, i.e. a lattice calculation. Since
the bottom quark is too heavy to be included directly in Monte Carlo
simulations on today's computers, the use of effective theories like HQET
is a necessity. In a recent
paper~\cite{KURTH1}, we tested HQET for a special correlation function, and
studied the discretisation errors connected with the heavy quark mass.

The operators under investigation are the axial current
\begin{equation}
  A_0(x)=\bar{\psi}_1(x)\gamma_0\gamma_5\psi_2(x)
\end{equation}
of two relativistic quark flavours $\psi_1$ and $\psi_2$,
and the static-light axial current
\begin{equation}
  A_0^{\rm stat}(x)=\bar{\psi}_1(x)\gamma_0\gamma_5\psi_{\rm h}(x),
\end{equation}
where the static quark field satisfies
\begin{equation}
  P_{+}\psi_{\rm h}=\psi_{\rm h},\qquad P_{+}=\frac12(1+\gamma_0),
\end{equation}
and its dynamics is
determined by the Eichten-Hill action~\cite{EICHHILL1}
\begin{equation}
  S_{\rm h}=a^4\sum_{x}\bar{\psi}_{\rm h}(x)\nabla_0^{\ast}\psi_{\rm h}(x),
\end{equation}

The correlation functions we use are defined in a finite space-time volume
$L\times L\times L\times L$ with Schr\"odinger functional boundary
conditions, i.e.
\begin{itemize}
\item The gauge field is periodic in space.
\item No spatial boundary conditions need to be specified for the static
quark field.
The relativistic quark fields are periodic in space up to a phase
$\theta$, e.g. (with unit vector $\hat{k}$)
\begin{equation*}
\psi_1(x+L\hat{k})=e^{i\theta}\psi_1(x).
\end{equation*}
\item Dirichlet boundary conditions in time.
\end{itemize}
The space-like link variables on the time boundaries are chosen as
$U_k(x)|_{x_0=0,L}=1$.
Correlation functions can contain derivatives $\zeta_i$,
$\bar{\zeta}_i$, $\zeta_i'$, $\bar{\zeta}_i'$, $\bar\zeta_{\rm h}$,
$\zeta_{\rm h}'$ with respect to the boundary quark fields
(see~\cite{LUSCHER1,KURTH2} for their definition).

We apply the Symanzik improvement programme~\cite{SYMANZIK1} to the axial
currents, defining improved operators $A_0^{\rm I}$ and $A_0^{\rm
stat,I}$, as well as to the action~\cite{LUSCHER1,KURTH2,MORNINGSTAR1}. 
We use the correlation functions
\begin{eqnarray}
  f_{\rm A}^{\rm I}(x_0)&=&-{{a^6}\over 2}\sum_{\bf y,z}
    \langle A_0^{\rm I}(x)
    \bar{\zeta}_2({\bf y})\gamma_5\zeta_1({\bf z})\rangle, \\
  f_1&=& -{{a^{12}}\over{2L^6}}\sum_{\bf u,v,y,z}
    \langle\bar{\zeta}_1'({\bf u})\gamma_5\zeta_2'({\bf v})\times
    \nonumber\\
    &&\times\bar{\zeta}_2({\bf y})\gamma_5\zeta_1({\bf z})\rangle, \\
  f_{\rm A}^{\rm stat,I}(x_0) & = & -{{a^6}\over 2}\sum_{\bf y,z}
    \langle A_0^{\rm stat,I}(x)\times\nonumber\\
    &&\times\bar{\zeta}_{\rm h}({\bf y})\gamma_5\zeta_1({\bf z})\rangle, \\
  f_1^{\rm stat} & = & -{{a^{12}}\over{2L^6}}\sum_{\bf u,v,y,z}
    \langle\bar{\zeta}_1'({\bf u})\gamma_5\zeta_{\rm h}'({\bf v})\times
    \nonumber\\
    &&\times\bar{\zeta}_{\rm h}({\bf y})\gamma_5\zeta_1({\bf z})\rangle.
\end{eqnarray}

\section{TESTING HQET}

Our aim is to compare HQET with full QCD \emph{in the continuum limit},
which requires renormalisation. A renormalised coupling is defined by
matching to the $\msbar$ scheme,
\begin{equation}
  g_{\msbar}^2=g_0^2+{\rm O}(g_0^4),
\end{equation}
where $g_0$ is the bare lattice coupling.
The quark masses are also renormalised by matching to the $\msbar$ scheme.
The $\msbar$ mass $m_{1,\msbar}$ of the first quark flavour is
chosen to be zero, while we introduce the dimensionless parameter
\begin{equation}
  z=Lm_{2,\msbar}(\mu=1/L)
\end{equation}
to parametrise the mass of the second quark flavour.
Taking the continuum limit means $a/L\rightarrow 0$ while keeping $z$
fixed. 
We form the ratios
\begin{eqnarray}
  X_{\rm I}(L/a)&=&
    {{f_{\rm A}^{\rm stat,I}(L/2)}\over{\sqrt{f_1^{\rm stat}}}},\\
  Y_{\rm I}(z,L/a)&=&{{f_{\rm A}^{\rm I}(L/2)}\over{\sqrt{f_1}}}.
\end{eqnarray}
in which the boundary field renormalisation constants and a divergence due
to the heavy quark self energy cancel. These ratios are expanded as
\begin{equation}
  \label{e_XYexpansion}
  X_{\rm I} = X_{\rm I}^{(0)}+X_{\rm I}^{(1)}g_0^2+{\rm O}(g_0^4),
\end{equation}
and analogously for $Y_{\rm I}$.
We introduce a
renormalisation scheme for the light-light axial current
by demanding that the renormalised ratio $Y_{\rm CA}$ satisfies the current
algebra relations at $z=0$. The static-light axial current requires a scale
dependent renormalisation,
and we define a renormalised ratio $X_{\rm match}(\mu)$ by
imposing the matching condition
\begin{eqnarray}
  X_{\rm match}(m_{2,\msbar})&=&Y_{\rm CA}(z,L/a)\nonumber\\
  &&+{\rm O}(1/z)+{\rm O}((a/L)^2).
\end{eqnarray}
The renormalised ratios are expanded in powers of the coupling analogous
to equation~(\ref{e_XYexpansion}).

\begin{figure}
  \vspace{3mm}
  \epsfig{file=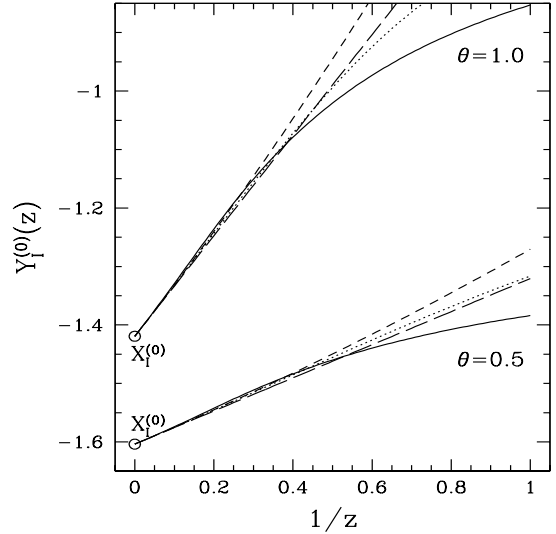,width=7.5cm,height=7.2cm,clip=}
  \vspace{-1.4cm}
  \caption{The ratio $Y_{\rm I}^{0}$ (solid curves) and the expansion of its
  analytic part at order $1/z$ (long dashes), $1/z^2$ (short dashes), and
  $1/z^3$ (dotted curves).}
  \vspace{-8mm}
  \label{f_TreeHQET}
\end{figure}

At tree level, the continuum limits $X_{\rm I}^{(0)}$ and 
$Y_{\rm I}^{(0)}(z)$ 
can be calculated analytically. $Y_{\rm I}^{(0)}(z)$
contains terms of the form $e^{-z}$, which are not analytic in
$1/z$. However, the analytic part can be expanded; the results at different
orders are shown in figure~\ref{f_TreeHQET}.
For $z>2$ the expansion
provides a good approximation, and the non-analytic part does not seem to
be significant.

\begin{figure}
  \epsfig{file=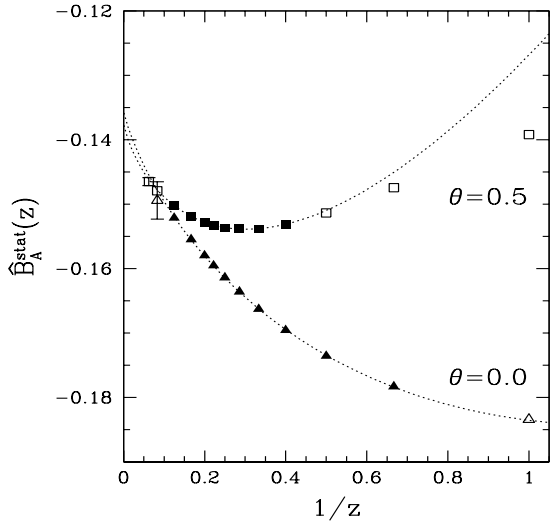,width=7.5cm,height=7.0cm,clip=}
  \vspace{-1.4cm}
  \caption{$\hat{B}_{\rm A}^{\rm stat}(z)$ in the continuum limit.
  The dotted curves are fits to the filled symbols,
  of the form predicted by HQET.}
  \vspace{-6mm}
  \label{f_OneLoopHQET}
\end{figure}

At one-loop level, HQET predicts that
\begin{eqnarray}
  X_{\rm match}^{(1)}(m_{2,\msbar})&=&X_{\rm I}^{(1)}(L/a)\nonumber\\
  &+&\left\{
  B_{\rm A}^{\rm stat}-\gamma_0\ln(z{a\over L})\right\}X_{\rm I}^{(0)}
  \nonumber\\
  &+&{\rm O}((a/L)^2),
\end{eqnarray}
with $\gamma_0=-1/4\pi^2$~\cite{VOLOSHIN1,POLITZER1}. 
This means that $B_{\rm A}^{\rm stat}$ can be
obtained as the $1/z\rightarrow 0$ limit of
\begin{eqnarray}
  \hat{B}_{\rm A}^{\rm stat}(z)&=&\gamma_0\ln(z{a\over L})
    \nonumber\\
    &+&{1\over{X_{\rm I}^{(0)}}}\left\{Y_{\rm CA}^{(1)}(z)-X_{\rm
    I}^{(1)}(L/a)\right\},
\end{eqnarray}
where it is understood that the continuum limit is taken on the right hand
side. 
$\hat{B}_{\rm A}^{\rm stat}$ 
is shown for two different values of the phase $\theta$ in 
figure~\ref{f_OneLoopHQET},
where fits of the expected form
\begin{equation}
  B_{\rm A}^{\rm stat}+\hat{f}_1{1\over z}+\hat{f}_2{1\over z}\ln(z)
  +{\rm O}(1/z^2)
\end{equation}
are also shown. 
The extrapolation results are
\begin{eqnarray}
  B_{\rm A}^{\rm stat}(\theta=0.0) & = & -0.136(3), \\
  B_{\rm A}^{\rm stat}(\theta=0.5) & = & -0.137(1).
\end{eqnarray}
This agrees with a previous result~\cite{BORRELLI1},
where matrix elements between quark states were used. 
Turning the argument around, this means that using the $B_{\rm
A}^{\rm stat}$ value from~\cite{BORRELLI1}, HQET describes the heavy quark
limit of QCD, independent from 
the chosen correlation functions. To our knowledge, this is the first
time that this fact has been established in an explicit example.

\section{DISCRETISATION ERRORS}

\begin{figure}
  \epsfig{file=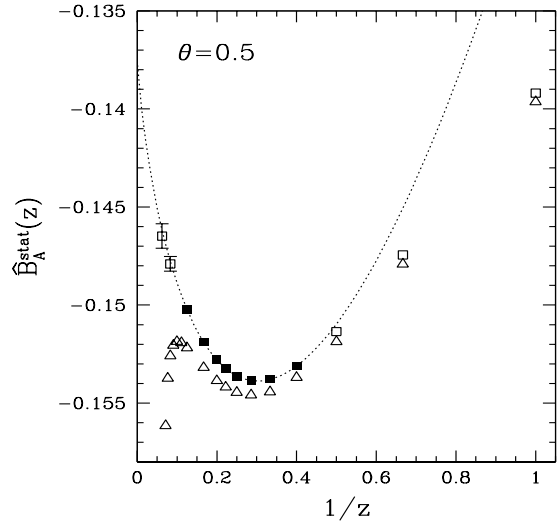,width=7.5cm, height=7.0cm,clip=}
  \vspace{-1.4cm}
  \caption{$\hat{B}_{\rm A}^{\rm stat}(z)$ in the continuum limit (squares)
  and for $L/a=32$ (triangles).}
  \vspace{-6mm}
  \label{f_DiscErr}
\end{figure}

A further interesting point to study is the quark
mass dependence of discretisation errors. We
discover that for not too large masses, the discretisation errors both at
tree-level and at one-loop order are
roughly proportional to $(am_{2,\msbar})^2$. However, for
$am_{2,\msbar}>1/4$, the behaviour of discretisation errors changes
completely. This is illustrated in figure~\ref{f_DiscErr}, 
where $\hat{B}_{\rm A}^{\rm
stat}$ is shown both in the continuum limit and at finite lattice spacing
$L/a=32$. 
A closer look shows
that this behaviour originates from a breakdown of ${\rm O}(a)$
improvement when the quark mass in lattice units becomes too large,
as it was previously observed in a perturbative analysis
of the Schr\"odinger functional coupling~\cite{SINT1}. This means that
extrapolations in the quark mass should be carried out in the continuum
limit rather than at finite lattice spacing.

This work is supported by the European Community's Human potential
programme under HPRN-CT-2000-00145 Hadrons/LatticeQCD.


\begin{thebibliography}{9}
\bibitem{KURTH1} M.~Kurth and R.~Sommer, hep-lat/0108018
\bibitem{EICHHILL1} E.~Eichten and B.~Hill,
                Phys. Lett. B 240 (1990) 193
\bibitem{LUSCHER1} M.~L\"uscher, S.~Sint, R.~Sommer and P.~Weisz,
                Nucl. Phys. B 478 (1996) 365
\bibitem{KURTH2} M.~Kurth and R.~Sommer, Nucl. Phys. B 597 (2001) 488
\bibitem{SYMANZIK1} K.~Symanzik, Nucl. Phys. B 190 (1981) 1
\bibitem{VOLOSHIN1} M.~B.~Voloshin and M.~A.~Shifman,
                Yad. Fiz. 45 (1987) 463
\bibitem{POLITZER1} H.~D.~Politzer and M.~B.~Wise, 
                Phys. Lett. B206 (1988) 681
\bibitem{MORNINGSTAR1} C.~Morningstar and J.~Shigemitsu,
               Phys. Rev D 57 (1998) 6741
\bibitem{BORRELLI1} A.~Borrelli and C.~Pittori,
                Nucl. Phys. B 385 (1992) 502
\bibitem{SINT1} S.~Sint and R.~Sommer, Nucl. Phys. B 465 (1996) 71
\end{thebibliography}
\end{document}